\documentclass[useAMS,usenatbib]{mn2e}
\usepackage{epsfig}
\usepackage{bethmacros}
\DeclareGraphicsExtensions{.eps, .jpg}

\def\spose#1{\hbox to 0pt{#1\hss}}
\def\lta{\mathrel{\spose{\lower 3pt\hbox{$\mathchar"218$}}
     \raise 2.0pt\hbox{$\mathchar"13C$}}}
\def\gta{\mathrel{\spose{\lower 3pt\hbox{$\mathchar"218$}}
     \raise 2.0pt\hbox{$\mathchar"13E$}}}

\def\inv{{${}^{-1}$}}

\title[Feedback and the Formation of Dwarf Halos]{Feedback and the Formation of Dwarf Galaxy Stellar Halos}

\author[Stinson et al.]{
{G.\,S. Stinson$^{1,2}$\thanks{Email: stinson `at' physics.mcmaster.ca},J.\,J. Dalcanton$^{1,4}$,T. Quinn$^{1}$,S. M. Gogarten$^{1}$,T. Kaufmann$^{3}$,J. Wadsley$^{3}$ 
}
\vspace*{6pt}\\
$^{1}$Astronomy Department, University of Washington, Box 351580, Seattle, WA, 98195-1580\\
$^{2}$Department of Physics and Astronomy, McMaster University, Hamilton, Ontario, L8S 4M1, Canada\\
$^{3}$Department of Physics and Astronomy, University of California, Irvine, Irvine, CA, 92697\\
$^{4}$Tom and Margo Wyckoff Fellow}

\begin{document}

\maketitle
\label{firstpage}

\begin{abstract} 
Stellar population studies show that low mass galaxies in all environments exhibit stellar halos that are older and more spherically distributed than the main body of the galaxy.  In some cases, there is a significant intermediate age component that extends beyond the young disk.  We examine
a suite of Smoothed Particle Hydrodynamic (SPH) simulations and find that elevated early star formation activity combined with supernova feedback can produce an extended stellar distribution that resembles these halos for model galaxies ranging from $v_{200}$ = 15 km s$^{-1}$ to 35 km s$^{-1}$, without the need for accretion of subhalos.  
\end{abstract}


\section{Introduction}
\label{intro}

The metal-poor stellar halo is the oldest extended structure in the
Galaxy \citep{oort24}.  Its internal structure, kinematics, and
metallicity encode information about the early assembly of the Galaxy.
Inspired by the constraints available in the Milky Way, several groups
have begun intensive observational studies of comparable structures in other large
spiral galaxies, most notably in M31 and M33
\citep{ferguson02,guhathakurta05,gilbert06,Ibata07}, in several dedicated HST
programs \citep[e.g., see preliminary results in][]{seth07,dejong07}, and in
isolated fields in nearby galaxies
\citep{harris00,mouhcine05a,Tikhonov2006,seth07,dejong2007}.  These studies use
resolved stellar populations to trace the extended structure of
galaxies down to surface brightnesses that are an order of magnitude
fainter than what is possible with ground-based surface photometry.

Both observations and simulations suggest that the stellar halos of massive galaxies are a
remnant of early hierarchical merging \citep{searle78}.  As galaxies
grow, they repeatedly accrete smaller galaxies.  Dynamical friction
pulls the accreted galaxies inwards, while tidal disruption strips the
galaxies' stars.  These stars are left on orbits
with large apocenters, creating an extended stellar halo.  Increasingly
mature numerical simulations now make specific predictions for the
profile of stellar halos, their degree of substructure, and the
metallicity of the accreted stars
\citep[cf.,][]{bekki03,brook05,bullock05,renda05,
bekki05,gauthier06,abadi06,Font06,Bekki2008}.

Less theoretical study \citep[cf.,][]{Mashchenko2008, Bekki2008,Valcke2008} has been devoted to the extended stellar structure of dwarf galaxies, in spite of the fact that old stellar populations have been observed surrounding every low mass galaxy that has been studied to sufficient depth.  We present a review of these observations in \S \ref{sec:haloobs}.  At first glance, the observations are consistent with dwarf halos being analogous to
the halos seen in the MW and other massive spirals, suggesting that
stellar halos may be a common by-product of galaxy assembly at all
mass scales.

In spite of their similarities, significant differences between dwarf stellar halos
and massive galaxy halos are becoming apparent, suggesting that they may not be true analogs of one another.\footnote{In spite of these differences, we will continue to refer to the low surface brightness, extended distribution of older stars surrounding dwarf galaxies as a ``halo" since it best describes the morphology of the extended stellar population.  However, the kinematics, metallicities, age distributions, and origins of the stars in the dwarf ``halo" are not necessarily exact analogs to the halo of the Milky Way.}  First,
deep color-magnitude diagrams (CMDs) constructed just outside the main optical body of the Phoenix ``transitional'' dSph \citep{Hidalgo2003b} and the Leo A dIrr \citep{Vanseviv2004} indicate a non-negligible population of
2-8\,Gyr old stars at these larger radii.\footnote{Such an
intermediate age population could not yet have been detected in other
dwarf galaxy CMDs, which are typically much shallower than the
observations of Phoenix and Leo A.}  This ``intermediate age'' population has no counterpart in the ancient halo of the Milky Way or the outer halo of M31 \citep{guhathakurta05, Kalirai2006}, although such a population does exist in the inner halo of M31, which is likely contaminated by thick disk stars \citep{brown03,Brown2008}.  Second, the surface density of dwarf
galaxies' stellar halos seem to fall off exponentially, rather than as
a power-law \citep[see radial
profiles in][ for example]{Vanseviv2004,Hidalgo2003b,Aparicio1997}.  The somewhat
brighter low mass galaxy NGC~4244 ($M_B=-17.0$) also appears to host a radially exponential distribution of halo stars, although a power law is not ruled out \citep{seth07}.

If we apply standard hierarchical structure formation arguments to form dwarf stellar halos, it is unclear how these features would be produced.  First, it is unlikely that any of the low mass sub halos that dwarf galaxies accreted contained  sufficient stellar material to populate the halos \citep{Purcell07}.  Second, since most merging happens early in the history of the Universe, it is unlikely that merging can routinely produce significant intermediate age stellar populations.

To understand how the mass-dependent differences in the properties
of stellar halos might arise, one should consider alternate possible origins
for the dwarf galaxy halos.  In this paper, we explore possible mechanisms for halo formation focusing on those that are intrinsic to the galaxies themselves rather than dependent on mergers or interactions.  

For isolated galaxies, stars located in an older stellar halo could have formed \emph{in situ} in a contracting envelope of star formation or could have formed in a central region and then been ejected.  Using our N-body simulations, we demonstrate that these mechanisms can indeed produce extended stellar halos in isolated dwarfs, without the need to accrete stars from merging subhalos.  We show that the resulting halos have exponential profiles and age gradients whose characteristic scale lengths agree with observations.  

This paper is structured as follows.  In \S \ref{sec:haloobs}, we present observational evidence for dwarf halos.  In \S \ref{sec:sims}, we describe the model we use for dwarf galaxies that form stellar halos.  In \S \ref{sec:results} we examine the age gradients, stellar density profiles, star formation histories, and halo shapes that are produced in the models.  

\section{Evidence for Dwarf Galaxy Halos}
\label{sec:haloobs}
Over the past decade, access to wide-field optical and near-infrared
imaging has increased dramatically.  These technological advances have
expanded the opportunities to obtain wide-field spatially-resolved
CMDs of the stars within dwarf galaxies.
Access to similar imaging capabilities in space, through the Hubble
Space Telescope, has allowed comparably high-quality CMDs to be
derived well beyond the confines of the Local Group, out to
$\sim\!5\Mpc$.

This growing body of observations has led to wide-spread confirmation
of what had already been suspected from ground-based surface
photometry, namely, that star-forming dwarf galaxies show
population gradients in their outer regions.  The CMDs of dwarf
irregular and ``transition'' dwarf galaxies universally show extended
envelopes dominated by older red giant branch (RGB) stars.  These
envelopes can extend many times further than the young bright blue
stars, and often have different axial ratios and morphologies than the
embedded young component or the extended distribution of HI
\citep[e.g., NGC~6822, IC~1613;][]{Battinelli2006a,Bernard2007}.

\begin{figure}
\resizebox{9cm}{!}{\includegraphics{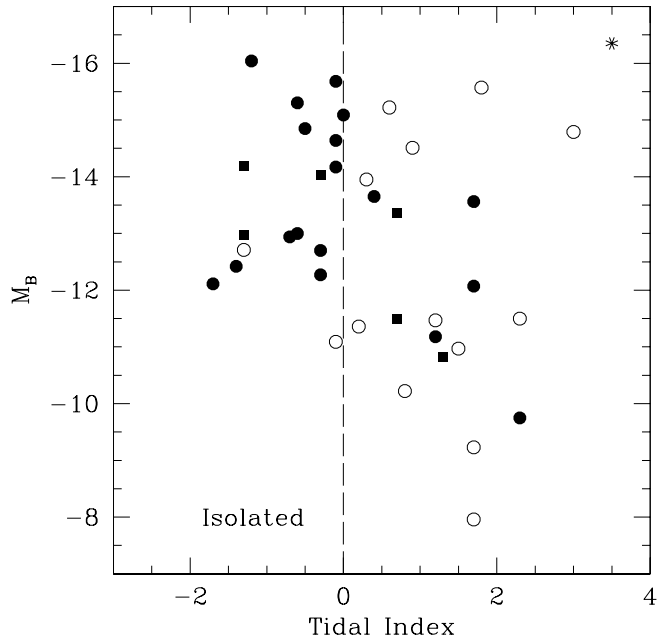}}
\caption[]{Absolute $B$-band magnitude and tidal index for dwarf
  galaxies with reported RGB halos.  Quantities are taken from
  \citet{karachentsev04}'s Catalog of Neighboring Galaxies.  The tidal
  index indicates the degree of expected tidal influence from nearby
  massive galaxies, with negative values indicating truly isolated
  systems.  Open circles are Local Group galaxies with the Milky Way
  or M31 as the primary perturber.  Solid squares are galaxies in the
  M81 group.  Solid circles are all other galaxies.  For reference,
  the asterisk in the upper right corner indicates the properties of
  the Small Magellanic Cloud.  }
\label{fig:obs} 
\end{figure}

A thorough review of the literature yields more than 50 papers on
$\sim\!40$ star forming dwarf galaxies hosting an extended halo of red
giant branch stars \citep{vanDyk1998, Alonso-Garcia2006, Aparicio1997, Aparicio2000, Aparicio2000a, Battaglia2006, Battinelli2004a, Battinelli2004, Battinelli2005, Battinelli2006, Battinelli2006a, Battinelli2007, Bernard2007, Caldwell1998, deJong2008, Demers2004, Demers2006a, Demers2006b, Doublier2000, Drozdovsky2000, Gallart2004, Han1997, Held2001, Held1999, Hidalgo2003a, hidalgo03, Hidalgo2003b, Hutchings1999, Lee1999, Lee1999a, Lee2005, Letarte2002, Martinez-Delgado1999, Martinez-Delgado1999a, McConnachie2006, Miller2001, Minniti2001, minniti03, Minniti1997, Minniti1999, Noel2007, Piersimoni1999, Rejkuba2000, Saviane2000, Schulte-Ladbeck1999, Schulte-Ladbeck2002, Tikhonov2005, Tikhonov2006, Tikhonov2006a, Vanseviv2004, Zijlstra1999}.  Earlier reviews can be found in
\citet{Saviane2001} and in part of \citet{Kunth2000}. There were no
cases reported in the literature where this effect was looked for, but
not found.\footnote{The one reported non-detection in NGC~3109 by
  \citet{Demers2003} is contradicted by \citet{Minniti1999}, who found
  an extended RGB envelope using data that reaches a full magnitude
  deeper.}  In Figure~\ref{fig:obs} we show the properties of the galaxies with extended RGB distributions as a
function of absolute $B$ magnitude and tidal index, both taken from
\citet{karachentsev04}'s Catalog of Neighboring Galaxies.  The tidal index
indicates the strength of expected tidal forces on the galaxy, based
on the nearest massive neighbors; negative values indicate galaxies
that are essentially isolated.

Figure~\ref{fig:obs} shows that the presence of an extended halo of
older stars is seen in all environments over a range of stellar masses.  The literature thus suggests that the formation process must be intrinsic to the galaxies themselves rather than due exclusively to tidal effects from larger galaxies.
Unfortunately, no data yet exist on the fraction of stellar mass in
the extended halo, and thus we can say only that such envelopes
exist, and are ubiquitous in low mass dwarfs.

Stellar population gradients also are found in the majority of dwarf
spheroidals with sufficiently extended observations \citep[Sculptor,
LeoII, AndI, AndII, NGC~185, NGC~147,
ESO~410-G005;][]{Tolstoy2004,Komiyama2007,
DaCosta1996,McConnachie2007,Martinez-Delgado1999,Han1997,Karachentsev2000}.
These observations are more subtle to interpret, due to the
lack of current star formation.  Instead, the population gradients typically manifest as
changes in horizontal branch morphology with radius, such that bluer,
more extended horizontal branches (favoring older ages and/or lower
metallicities) are seen towards the outskirts.  The gradients can also
be traced using intermediate age asymptotic giant branch (AGB) stars,
which are found to be more centrally concentrated.  Unlike the dwarf
irregulars and transition galaxies, however, there are 4 well-studied
dwarf spheroidals (LeoI, Carina, Tucana, and possibly Draco) in which
gradients have not been detected
\citep{Held2000,Saviane1996,DaCosta2000,Segall2007}.  However, given
that the likely formation scenarios for dwarf spheroidals involve
extensive dynamical threshing of dwarf irregulars as they are accreted
by larger halos \citep[e.g.,][]{Mayer2007}, it is more surprising that
any dwarf spheroidals could have retained population gradients imprinted
before they fell in.

While the evidence from color-magnitude diagrams confirms the presence
of older stars in the envelopes of dwarf galaxies, few observations
yet constrain exactly how old those stars are.  A simple stellar
population forms a prominent red giant branch after only a Gigayear or
so, and thus some of the stars in the RGB envelopes might be of
intermediate age, provided that the main sequence turnoff is too faint
to be detected in the existing observations.  Observations of galaxies
beyond the Local Group rarely reach the horizontal branch, and
therefore are incapable of detecting main sequence turnoffs of younger
than $\sim\!1 \rm{Gyr}$.  The RGB envelopes may therefore host an
intermediate age stellar population, and may not necessarily be strict
analogs of the ancient stellar halo of the Milky Way.  Indeed,
extensive near-infrared observations of AGB stars in the halos of
dwarf galaxies demonstrates the presence of intermediate age
(1-10$\rm{Gyr}$) populations
\citep[e.g.,][]{Demers2004,Battinelli2006,Battinelli2004a,
Battinelli2004,Demers2006a,Letarte2002}.  Such intermediate age
populations have also been verified by direct detection of main
sequence turnoffs in the transition dwarf Phoenix by
\citet{hidalgo03}, although the fraction of intermediate age stars
is small.  Given the limited evidence to date, one should not assume
that the ``halos'' of red giant branch stars around dwarfs are purely
ancient.

Taken together, the extensive body of literature above indicates that
extended envelopes of red giant branch stars and strong radial
population gradients are common in dwarf galaxies, with the exception
of some subset of dwarf spheroidals in the Local Group.  The gradients
indicate that the outer envelopes are older than the central galaxy,
but that they may contain some fraction of intermediate age stars.
The ubiquity of the phenomena in galaxies of all environments
indicates that halos can be produced by phenomena internal to the
galaxy, rather than being induced by tidal interactions with massive
companions.  The presence of intermediate age stars in some halos
further indicates that dwarf halos are not necessarily a by-product of
processes that operate only during the initial formation of the galaxy.

\section{Simulations}
\label{sec:sims}
To examine dwarf halo formation, we begin using the same suite of simulations described in \citet{Stinson07}.  The model galaxies vary in total mass from $10^9 {\rm M}_\odot$ to $10^{10} {\rm M}_\odot$ and have velocities at the virial radius ($v_{200}$) of 15, 20, 25, and 30 km s$^{-1}$, resulting in stellar velocity dispersions comparable to Local Group dwarfs.  Each model galaxy initially consists of a sphere of $10^5$ gas particles in hydrostatic equilibrium with a sphere of $10^5$ dark matter particles, initially distributed with a \citet{NFW} density profile ($c = 8$).  The baryons account for 10\% of the total galaxy mass.  Dark matter particles in every model are left with their random equilibrium velocities \citep{Stelios04} while the gas is spun with a uniform circular velocity so that $\lambda=\frac{j_{gas}\left|E\right|^{\frac{1}{2}}}{GM^{\frac{3}{2}}}$ = 0.039, where $j_{gas}$ is the average specific angular momentum of the gas, and $E$ and $M$ are the total energy and mass of the halo.  Although these initial conditions are not ideal analogs of the likely mechanisms for gas accretion, the mechanisms we explore in this paper only manifest after the formation of a settled star forming system.  Thus, the properties of the resulting halos are unlikely to depend sensitively on transient effects linked to the specific choice of initial conditions.

We supplement these models with a higher resolution simulation of the 15 km s$^{-1}$ galaxy
to maximize the number of particles in the diffuse
stellar halo.  The high resolution model is the same in every respect to those described above except that it contains ten times more gas and dark matter particles.  Each gas particle in the high resolution simulation initially contains 140 M$_\odot$ of material, and each star particle represents an initial mass in stars of 40 M$_\odot$.  

To these, we add simulations with varying baryon fraction.  In a cosmological context, reionization has a strong impact on small galaxies and puts an effective mass limit on galaxies that can retain enough gas to form stars \citep{Quinn96, HaardtMadau, BarkanaLoeb99, Bullock00}.  We have not included reionization explicitly, and instead capture the relevant physics by simply excluding increasing fractions of gas from the galaxies as would be expected if a UV background prevented some fraction of the gas from cooling into a disk.  A more realistic simulation that included the full cosmological context would be required to verify this scenario, but
resolving structure at this scale would be prohibitively expensive (cf. \citet{Mashchenko2008} require $10^6$ processor hours to evolve such a dwarf to $z = 5$).  Empirically, low mass galaxies have high mass to light ratios, and at least an order of magnitude scatter in their inferred baryon fraction \citep{vdB2003, Lin2004, Yang2005, Hoekstra2005, Mandelbaum2006, Zaritsky2007}.  To mimic this, we simulate three 35 km s$^{-1}$ galaxies with baryon fractions ($f_b$) of 10 \%, 1\%, and 0.5\%.  The varying $f_b$ galaxies contain $10^5$ gas and $10^5$ dark matter particles, but all have a total mass of $1.365\times10^{10} {\rm M}_\odot$, the mass above which \citet{Quinn96} find a galaxy can survive reionization ($v_{200}$=35 km s$^{-1}$).  

The simulations were run using the parallel SPH code \textsc{gasoline} \citep{gasoline}.  \textsc{gasoline} solves the equations of hydrodynamics, and includes radiative cooling.  The cooling assumes ionization equilibrium, an ideal gas with primordial composition (metal line and H$_2$ cooling are not included), and solves for the equilibrium abundances of each ion species.  The scheme uses the collisional ionization rates reported in \citet{Abel97}, the radiative recombination rates from \citet{Black81} and \citet{Verner96}, bremsstrahlung, and line cooling from \citet{Cen92}. The energy integration uses a semi-implicit stiff integrator independently for each particle with the compressive heating and density
(i.e. terms dependent on other particles) assumed to be constant over the
timestep.  Gravity is calculated for each particle using tree elements that span
at most $\theta$ = 0.7 of the size of the tree element's distance from
the particle.  \textsc{gasoline} is multistepping so
that every particle's time-step
$\Delta{t_{\rm grav}}=\eta\sqrt{\frac{\epsilon_i}{a_i}}$, where $\eta$ =
0.175, $\epsilon_i$ is the gravitational softening length, and $a_i$
is the acceleration.  For gas particles, the time-step must also be
less than $\Delta{t_{\rm gas}}=\eta_{\rm Courant}\frac{h_i}{(1+\alpha)c_i + \beta\mu_{MAX}}$, where
$\eta_{\rm Courant}$ = 0.4, $h_i$ is the gas smoothing length, $c_i$ is the sound speed, $\alpha=1$ is the shear coefficient, $\beta=2$ is the viscosity coefficient, and $\mu_{MAX}$ is described in full detail in \citet{gasoline}. 

The star formation and feedback recipes are the ``blastwave model" described in detail in \citet{Stinson06}, and they are summarized as follows.  Gas particles must be dense ($n_{\rm min}=0.1 cm^{-3}$) and cool ($T_{\rm max}$ = 15,000 K) to form stars.  A subset of the particles that pass these criteria are randomly selected to form stars based on the commonly used star formation equation, 
\begin{equation}
\frac{dM_{\star}}{dt} = c^{\star} \frac{M_{gas}}{t_{dyn}}
\end{equation}
where $M_{\star}$ is mass of stars created, $c^{\star}$ is a constant star formation efficiency factor, $M_{gas}$ is the mass of gas creating the star, $dt$ is how often star formation is calculated (1 Myr in all of the simulations described in this paper) and $t_{dyn}$ is the gas dynamical time.  The constant parameter, $c^{\star}$ is tuned to 0.05 so that the simulated Isolated Model Milky Way used in \citet{Stinson06} matches the \citet{Kenn98} Schmidt Law, and then $c^\star$ is left fixed for all subsequent applications of the code.  This star formation and feedback treatment was one of the keys to the success of \citet{Governato07} in producing realistic spiral galaxies in a cosmological simulation and the success of \citet{Brooks07} in matching the observed mass-metallicity relationship.

At the resolution of these simulations, each star particle represents a large group of stars (40-3500 M$_\odot$).  Thus, each particle has its stars partitioned into mass bins based on the initial mass function presented in \citet{Kroupa93}.  These masses are correlated to stellar lifetimes as described in \citet{Raiteri96}.  Stars larger than 8 $M_\odot$ explode as supernovae during the timestep that overlaps their stellar lifetime after their birth time.  The explosion of these stars is treated using the analytic model for blastwaves presented in \citet{MO77} as described in detail in \citet{Stinson06}.  While the blast radius is calculated using the full energy output of the supernova, less than half of that energy is transferred to the surrounding ISM, $E_{SN}=4\times10^{50}$ ergs.  The rest of the supernova energy is radiated away.

\section{Results}
\label{sec:results}

In this section, we examine the spinning gas halos that cool radiatively into centrifugally and pressure supported star forming disks.  First, we provide evidence for their extended exponential stellar density profile and the presence of an age gradient in the models.  We then examine the mechanisms responsible for these features.

\subsection{Density Profiles}

\begin{figure}
\resizebox{9cm}{!}{\includegraphics{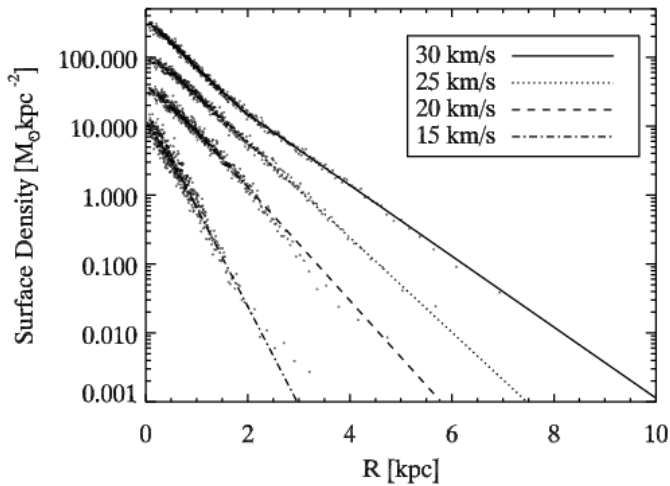}}
 \caption[Stellar Density Profiles]{ Projected edge-on radial stellar density profiles of the simulated galaxies.  Stars are averaged over radial bins that each contain equal numbers of star particles.  The lines correspond to the best two piece exponential fits to each radial set of points.  The more massive halos have higher central surface densities and longer exponential scale lengths.  Note that these plots do not show the number of stars brighter than some magnitude limit (i.e. Figure \ref{fig:cmds}), and thus they are not exact analogs of observational studies.  }
\label{fig:starprof} 
\end{figure}

In Figure \ref{fig:starprof} we show the radial surface density profile of the stars formed
in the simulations for several different mass simulations.  The halo surface density profiles generally follow an exponential, though there are breaks where the scale lengths change.  We fit these profiles with a double exponential.  The inner and outer exponential scale lengths are reported in Table \ref{tab:data} along with the break radius that separates each exponential segment in Figure \ref{fig:starprof}.  Similar broken exponential profiles are evident in observations; for example, see Figure 1 in \citet{hidalgo03} and Figure 3 in \citet{Vanseviv2004}.  Alternatively, the surface density profiles can be fit with a series of power laws, with a flat inner core, a middle region that declines as $r^{-2}$, and an outer region that declines steeply as $r^{-6}$ or steeper.  Although the two component exponentials are reasonable fits to the overall profile, there are significant deviations that are more pronounced in the lower mass galaxies.  If mechanisms involving supernova feedback create these outer halos, it makes sense that deviations would be more pronounced in shallower potential wells where the supernovae have a larger relative effect on the gas dynamics.  Figure \ref{fig:starprof} also shows that the stellar scale length increases with mass, as expected from analytic models \citep{Dalcanton97,mo98,vdB2001}.

\begin{table*}
\caption{Simulation data}
\begin{center}
\begin{tabular}{c|c|c|c|c|c|c|c|c}
M$_{tot}$ &baryon &v$_{200}$ &$\sigma_v$ &M$_\star$ &h$_{in}$ &h$_{out}$&r$_{break}$&r$_{Holm}$\\
(M$_\odot$)&fraction &(km s$^{-1}$)&(km s$^{-1}$)&(M$_\odot$)&(kpc)&(kpc)&(kpc)&(kpc)\\
\hline
$10^9$&10\%&15&7.4&$2.34\times10^6$&0.37&0.30&0.71&0.80\\
$2.5 \times 10^9$&10\%&20&11.4&$2.15\times10^7$&0.67&0.53&0.90&1.24\\
$5 \times 10^9$&10\%&25&15.1&$7.86\times10^7$&0.97&0.63&0.54&1.33\\
$8.6 \times 10^9$&10\%&30&20.1&$2.20\times10^8$&0.63&0.84&1.95&1.24\\
$1.4 \times 10^{10}$&0.5\%&35&15.6&$5.23\times10^6$&0.65&0.47&2.09&1.06\\
$1.4 \times 10^{10}$&1\%&35&16.8&$1.72\times10^7$&0.84&0.61&2.37&1.15\\
$1.4 \times 10^{10}$&10\%&35&29.9&$3.86\times10^8$&0.35&0.91&3.45&1.24\\
\end{tabular}
\end{center}
\label{tab:data}
\end{table*}

\begin{figure}
\resizebox{9cm}{!}{\includegraphics{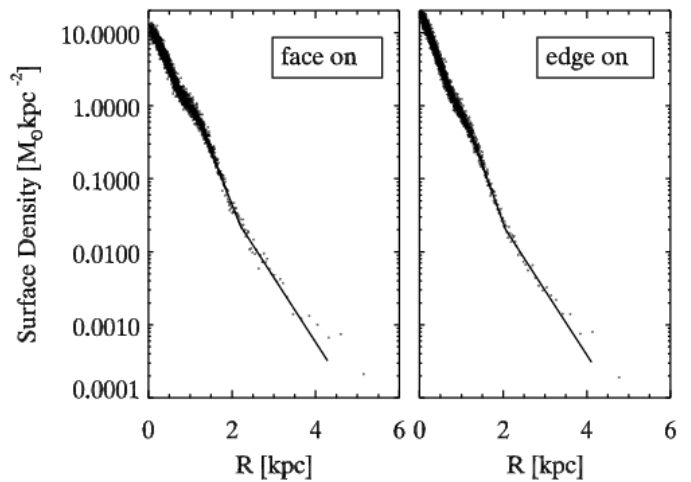}}
 \caption[High resolution surface density profile]{ Projected face-on and edge-on radial stellar density profiles of the high resolution 15 km s\inv model.  Stars are averaged over radial bins that each contain equal numbers of star particles.  The lines correspond to the best quadruple piece exponential fit to each radial set of points.  In the face on profile, the segments ending at 0.49, 0.8, 1.22, 2.22, and 4.5 kpc have scale lengths of 0.35, 0.52, 0.30, 0.49, and 0.81 kpc, respectively.  The edge-on view segments ending at 0.49, 0.8, 1.22, 2.08, and 4.5 kpc have scale lengths of 0.29, 0.41, 0.27, 0.49, and 0.82 kpc.}
\label{fig:hrstarprof} 
\end{figure}

Figure \ref{fig:hrstarprof} shows face-on and edge-on projections of the surface density profile of the high resolution 15 km s $^{-1}$ model.  The high resolution profile can be modeled with a series of exponential segments with scale lengths ranging from 0.3 to 0.8 kpc for both projections.  The outer scale lengths for the two orientations become more similar at large radii suggesting an increasingly isotropic, spherical shape with radius.  We return to this point in \S \ref{sec:bdps}.

\subsection{Age Gradients}

\begin{figure}
\resizebox{9cm}{!}{\includegraphics{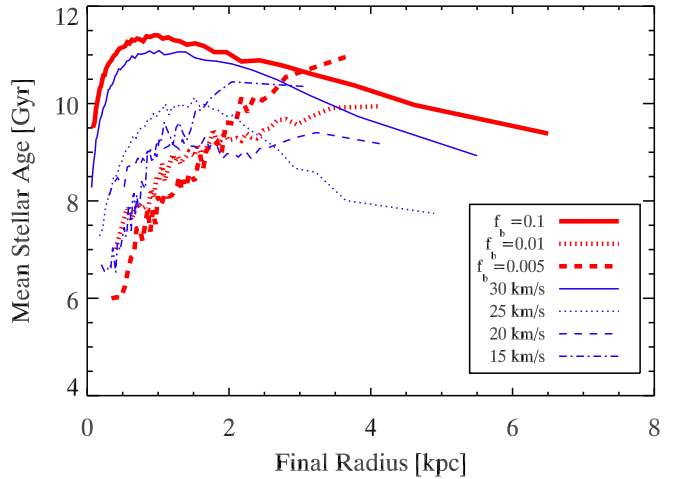}}
 \caption[Age Profile]{ The mean stellar age versus radius.  The mean is taken of stars grouped together into 40 radial bins with equal numbers of particles.  The thin lines represent the models run with a baryon fraction of 10\% and variable mass.  The thick lines represent the models that have a constant mass corresponding to $v_{vir}$ = 35 km s$^{-1}$ and variable baryon fraction.}
\label{fig:ageprofile} 
\end{figure}

\begin{figure*}
\resizebox{18cm}{!}{\includegraphics{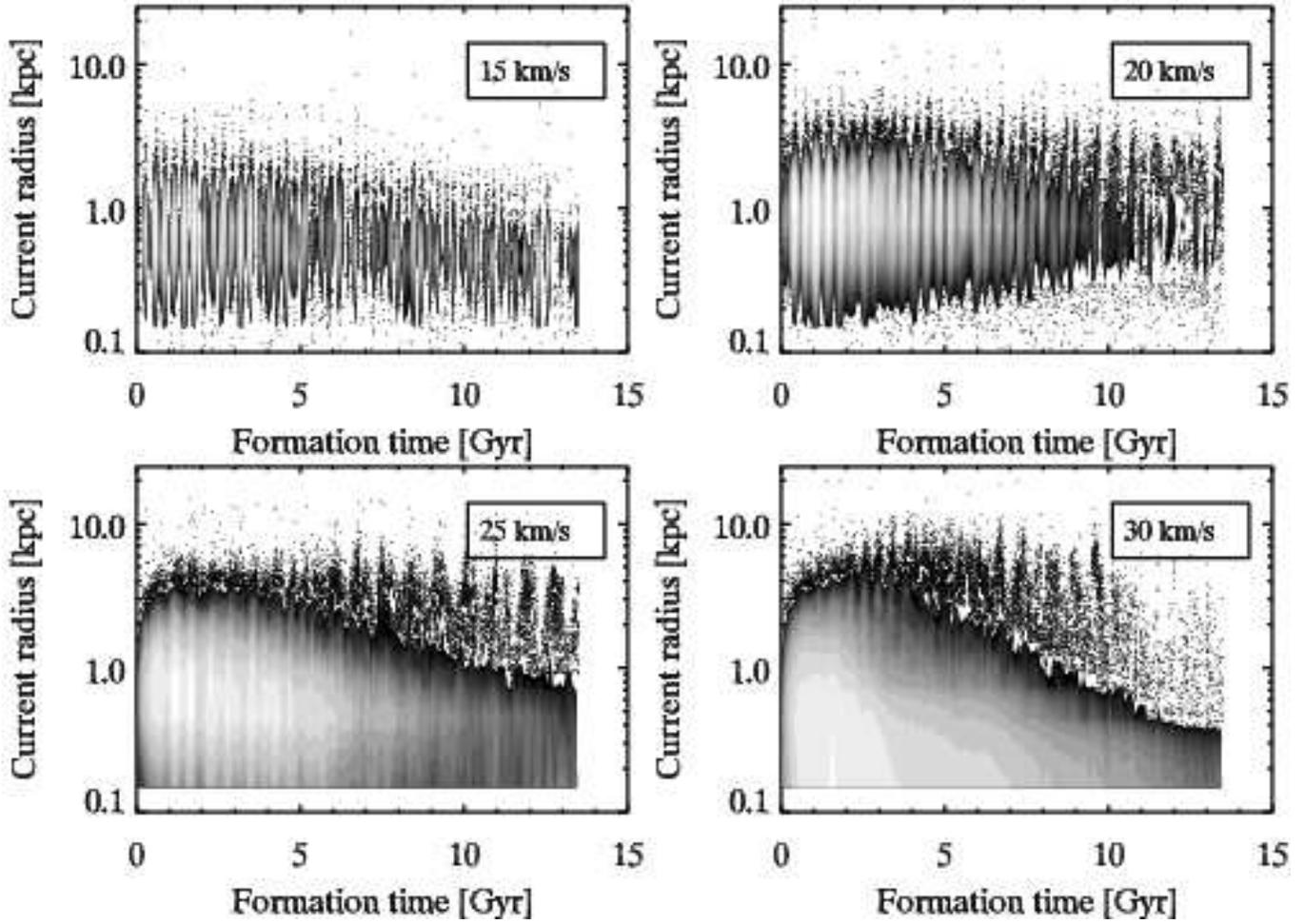}}
 \caption[Formation Time vs. Radius]{ The final distribution of where stars are  versus their formation time for the four models with $f_b$=10\% and varying masses represented by their $v_{200}$ velocities.  The plot replaces clusters of dots with contours in particularly dense regions.  The contours are are based on a density map with cells that are 0.3 kpc by 0.1 Gyr.  The contours are scaled between 10 levels from a minimum of 20 stars per cell (\emph{black}) to the maximum of stars in a given cell (\emph{white} inside \emph{black}).  Younger ages are to the right. }
\label{fig:kmsrtform} 
\end{figure*}

\begin{figure*}
\resizebox{18cm}{!}{\includegraphics{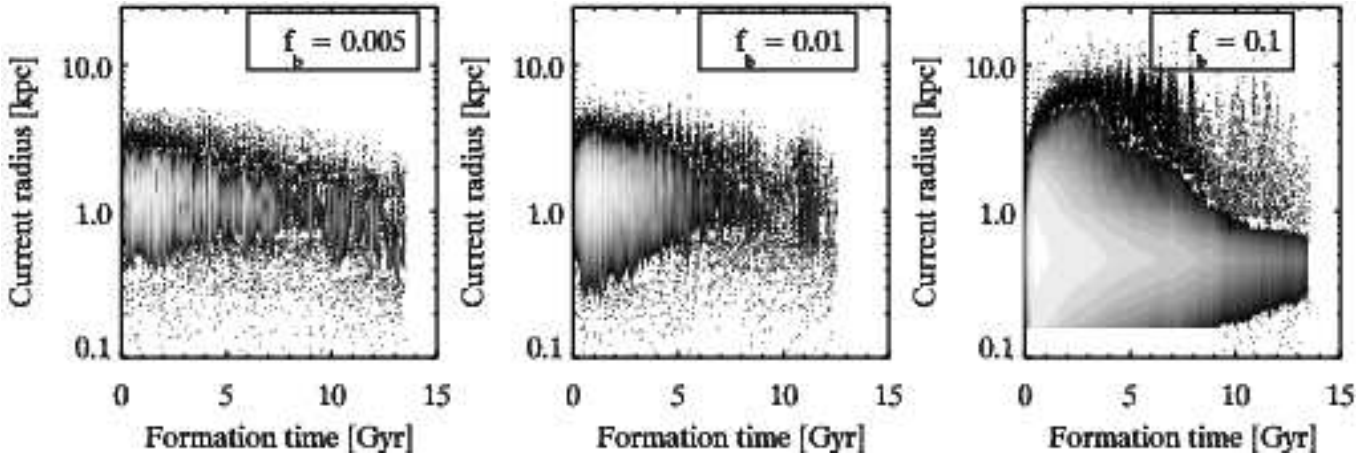}}
 \caption[Formation Time vs. Radius]{ The same as Figure \ref{fig:kmsrtform} except for the models with $v_{200}$ = 35 km s$^{-1}$ and $f_b$ = 0.005, 0.01, and 0.1.}
\label{fig:bfrtform} 
\end{figure*}

\begin{figure*}
\resizebox{18cm}{!}{\includegraphics{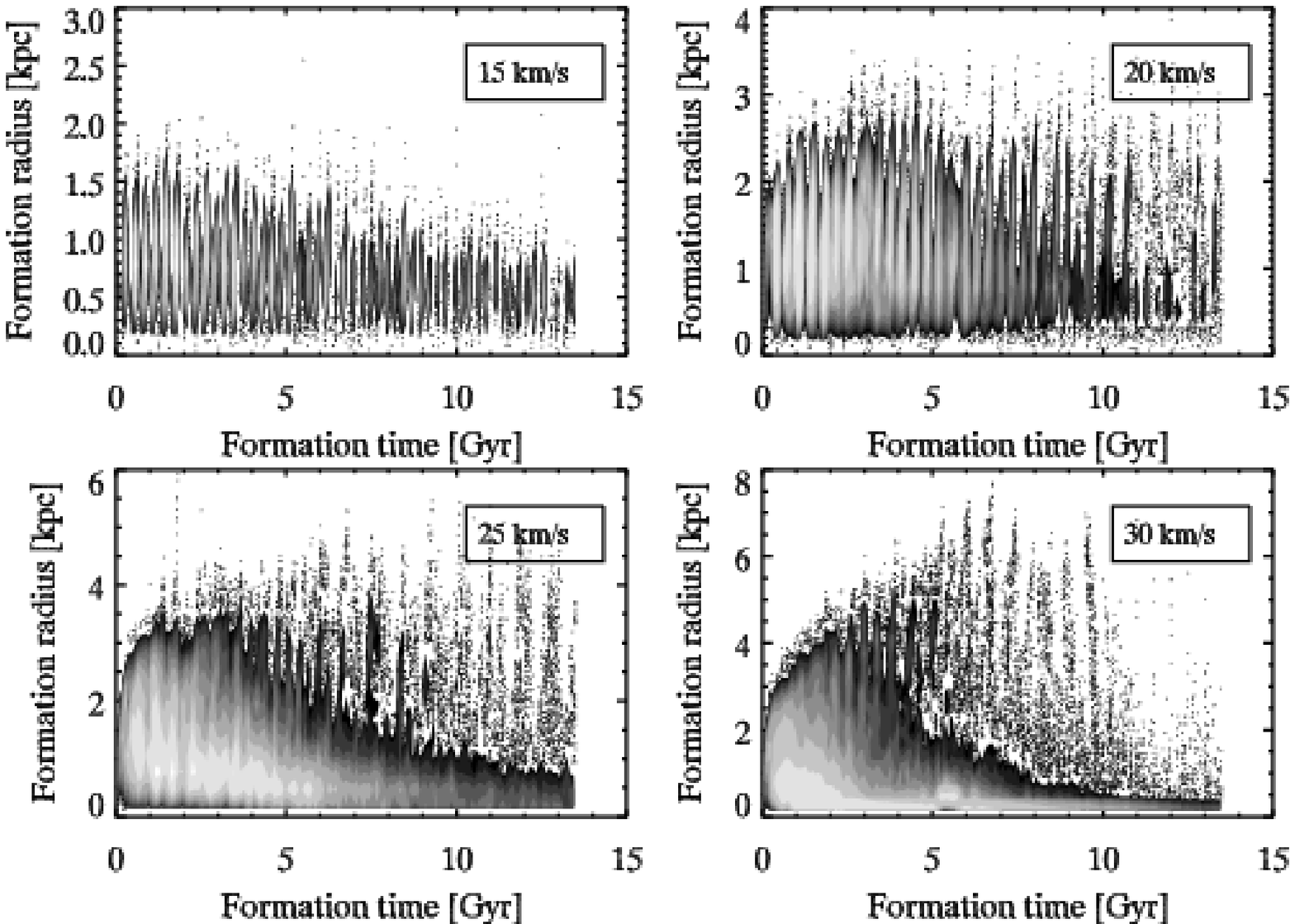}}
 \caption[Formation Time vs. Radius]{ The distribution of where stars are formed versus when they formed in the four models with $f_b$=10\% and varying masses represented by their $v_{200}$ velocities.  The contours are are based on a density map with grids that are 0.3 kpc by 0.1 Gyr.  The contours are scaled between 10 linearly spaced levels from a minimum of 20 stars per cell to the maximum of stars in a given cell.}
\label{fig:kmsrformtform} 
\end{figure*}

\begin{figure*}
\resizebox{18cm}{!}{\includegraphics{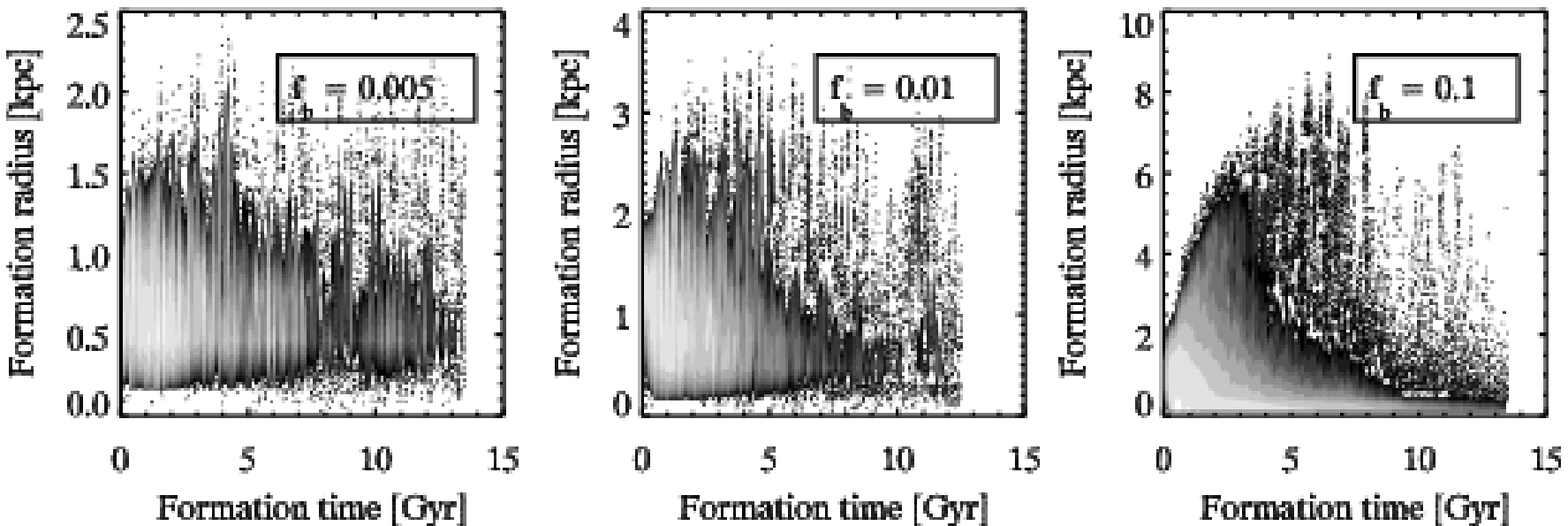}}
 \caption[Formation Time vs. Radius]{ The same as Figure \ref{fig:kmsrformtform} except for the models with $v_{200}$ = 35 km s$^{-1}$ and $f_b$ = 0.005, 0.01, and 0.1.}
\label{fig:bfrformtform} 
\end{figure*}

\begin{figure*}
\resizebox{18cm}{!}{\includegraphics{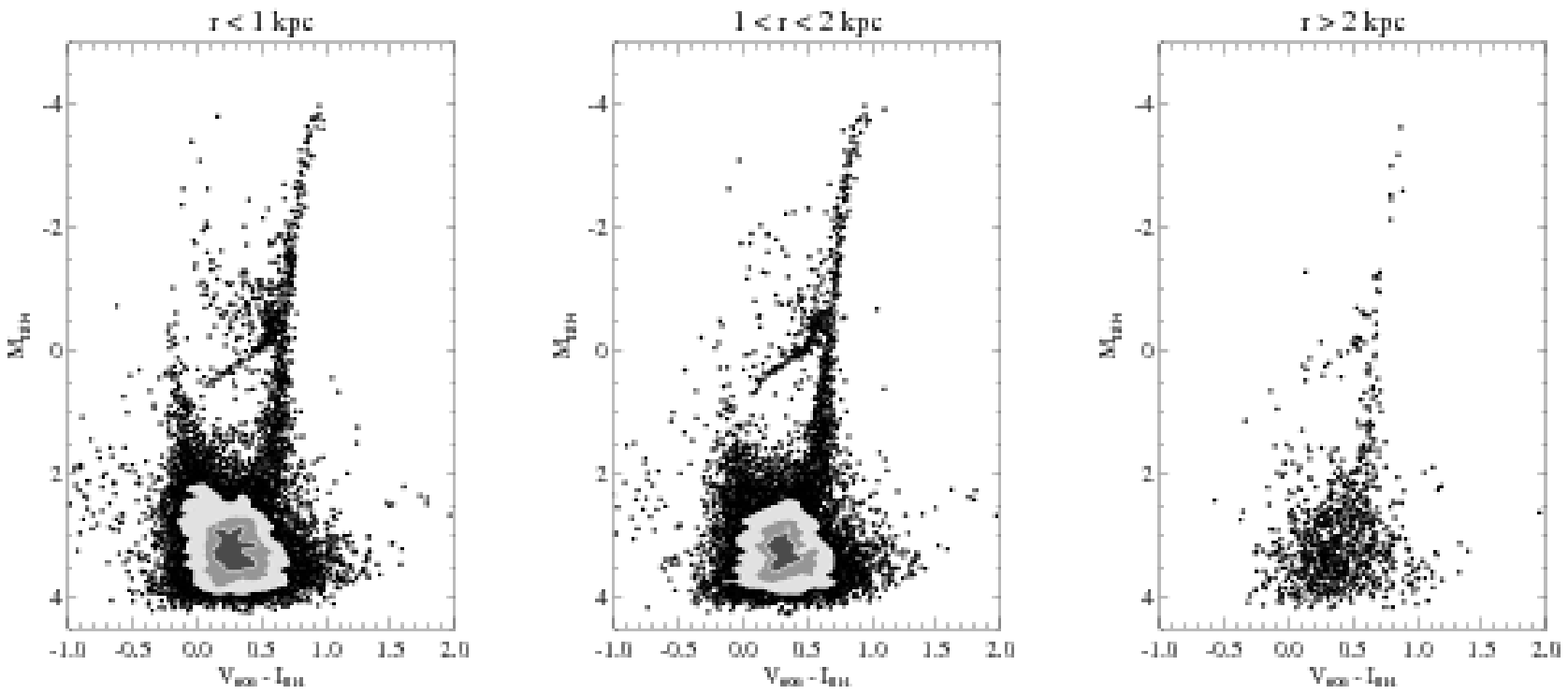}}
  \caption[Synthetic color magnitude diagrams]{Synthetic color magnitude diagrams of the 15 km s$^{-1}$, $f_b$=0.1
  simulated dwarf galaxy, shown in three radial bins.  Magnitudes are
  in the HST/ACS filter system.  The contour colors, from light to
  dark, represent point densities of 30, 50, and 70 points per $0.1
  \times 0.1$ mag bin.}
  \label{fig:cmds}
\end{figure*}

CMDs of dwarf galaxy halos show well populated red giant branches with little evidence for bright main sequence stars (\S \ref{sec:haloobs}).  These data suggest that the halos are systematically older than the central star forming regions.  We identify a similar age gradient by examining the distribution of stars in our models three different ways.  We first show the mean age gradient (Figure \ref{fig:ageprofile}).  We then look in detail at the full distribution of stellar ages as a function of position (Figures \ref{fig:kmsrtform}, \ref{fig:bfrtform}, \ref{fig:kmsrformtform}, and \ref{fig:bfrformtform}). Finally, we generate model CMDs from our simulations within radial bins (Figure \ref{fig:cmds}).  

Figure \ref{fig:ageprofile} shows the mean age profile of stars as a function of radius.  In all the halos, stars at the smallest radii have the youngest mean ages.  However, the mean age in the center of the galaxies is never less than 6 Gyr, the mean age expected if stars have been forming continuously for 12 Gyr.  In the lowest stellar mass halos (15 km s$^{-1}$ and $f_b$=0.005), the stars at large radii have the oldest mean ages.  In high stellar mass halos, the age profile turns over to slightly younger ages at larger radii, as might be expected for classic inside-out disk growth.  Lower mass galaxies do not show this trend because they lack a stable star forming disk.  In all cases, however, the outer regions are quite old ($>$ 8 Gyr).

Figures \ref{fig:kmsrtform} and \ref{fig:bfrtform} provide a more detailed examination of the behavior seen in Figure \ref{fig:ageprofile}.  The plots show the final location of stars versus their formation time.  They reveal that the disk evolution is more complicated than the age profiles indicate.  In all cases, there are few truly young stars at large radii; beyond 2 kpc, many fewer stars form after 5 Gyr.  Thus, in contrast to the classical ``inside-out'' formation expected for massive disks, the characteristic radius for star formation shrinks with time for these smaller systems.  Our high mass model galaxies do show an early period of ``inside-out'' growth during which the small age gradient seen in Figure \ref{fig:ageprofile} is imprinted at large radii.  However, after 5 Gyr, they too experience the same contraction in the star forming region of the disk that is seen in lower mass galaxies.

The contracting star formation results from the interaction between gas pressure and the star formation density criterion.  The gas disks of these small systems are partially supported by gas pressure in addition to classical angular momentum support \citep{Rhee2004,Valenzuela2007,Kaufmann2007}.  Therefore, as gas is consumed in the center due to initial high star formation rates, the gas disk contracts radially until pressure support is re-established.  The resulting decline in central star formation rate may also reduce the turbulent velocity of the gas, further reducing the pressure support.  This process leads to radial shrinking of the gas disk such that high gas densities are maintained in the center, at the expense of the outer regions.  The resulting evolution leads to star formation that is systematically less radially extended as time goes on.

We can verify that the age gradient observed in our simulations is established at formation rather than through stellar migration.  Figures \ref{fig:kmsrformtform} and \ref{fig:bfrformtform} show the radius where stars formed as a function of time.  The overall shape of Figures \ref{fig:kmsrformtform} and \ref{fig:bfrformtform} closely resembles Figures \ref{fig:kmsrtform} and \ref{fig:bfrtform}; both sets of figures show shrinking of the radial envelope within which stars form.  Formation location is therefore a primary factor in determining the final age gradient, rather than the degree of stellar migration.  One aspect of Figures \ref{fig:kmsrtform} and \ref{fig:bfrtform} that differs greatly from Figures \ref{fig:kmsrformtform} and \ref{fig:bfrformtform} is that many stars finish the simulation beyond where they form.  This outward movement is investigated in more detail in the next section. 

Each of the models shows two distinct spatial regions.  At small radii, the star formation is abundant and nearly continuous.  At large radii, there is less star formation and what there is is episodic.  During these episodes of elevated star formation, gas temporarily reaches star forming densities in the outer region of the disk, typically in spiral patterns.  Once the stars have formed, the supernova feedback disperses the dense gas and shuts down star formation.  

In models with lower stellar mass, the episodic bursts are scattered at regular time intervals spaced 300 Myr apart.  This timescale is similar to the star formation interval reported in \citet{Stinson07}, and is related to the free fall time at the center of halos.  Higher mass galaxies also show similar episodic star formation, but only outside the inner stable star forming disk.  At large radii, the episodic star formation has a longer timescale of $\sim 1$ Gyr due to the lower characteristic densities associated with longer free fall times.  This episodic star formation can be explained with a delay differential equation as shown by \citet{Quillen2008}.

For a more direct comparison with observations, Figure~\ref{fig:cmds} plots model CMDs showing the expected
stellar populations in several radial bins.  To create the CMDs, we
used StarFISH version
1.1 \citep{harris01} to populate a set of isochrones \citep{Girardi02} based on a
user-supplied star-formation history (SFH) set by the mass, age, metallicity (-2.3 $<$ [Fe/H] $<$-1 after 13.5 Gyr), and
position of the simulated star particles.
To mimic observing conditions, we adopted artificial star tests from the ACS Nearby Galaxy Survey \citep{Dalcanton2008} assuming that the simulated galaxy is
located at a distance of 500 kpc.

The CMDs of the simulations show a prominent young main sequence in the central
 2 kpc that disappears at large radii.  The absence of the young main sequence
outside 2 kpc agrees with the age gradient shown in Figure \ref{fig:ageprofile}.  Figure \ref{fig:cmds} also shows that any main sequence stars in the halo
 are more than 1 magnitude fainter than the red clump and horizontal
branch, such that only the deepest photometry in the nearest galaxies would show evidence of this intermediate age population.  The age gradient apparent in Figure \ref{fig:cmds} is consistent with the deep CMDs presented in Figure 2 of \citet{hidalgo03} suggesting that isolated models can produce realistic stellar structure.

\subsection{Stellar Migration}
\label{sec:movement}
\begin{figure*}
\resizebox{18cm}{!}{\includegraphics{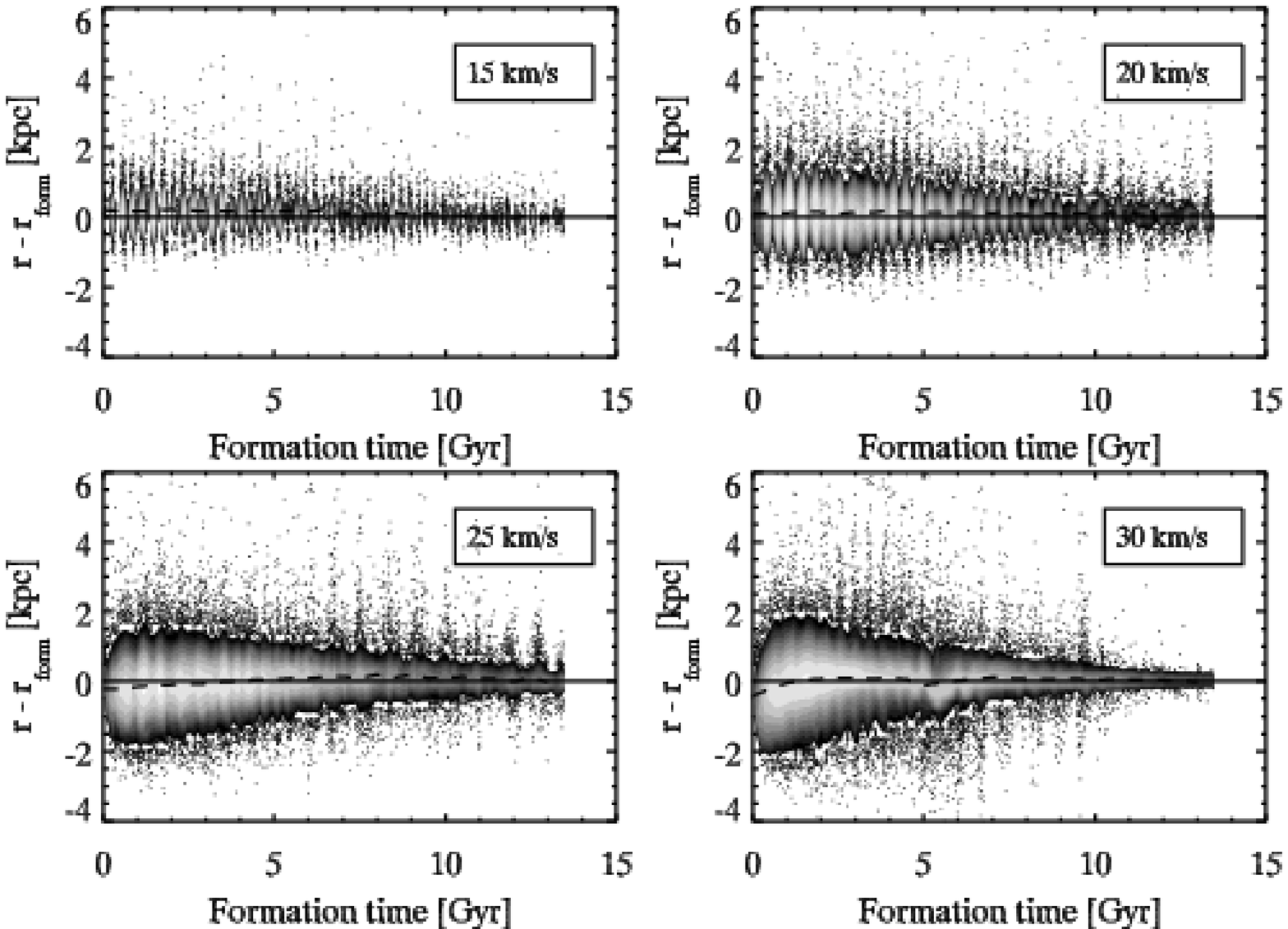}}
 \caption[Formation Time vs. Movement]{ The distribution of how much stars moved versus when they formed in the four models with $f_b$=10\% and varying masses represented by their $v_{200}$ velocities.  The contours are are based on a density map with grids that are 0.3 kpc by 0.1 Gyr.  The contours are scaled between 10 linearly spaced levels from a minimum of 20 stars per cell to the maximum of stars in a given cell.  The dashed line represents a lack of movement from the stars formation location.  The dotted line represents the mean movement of particles at the various formation times.}
\label{fig:kmsmigration} 
\end{figure*}

\begin{figure*}
\resizebox{18cm}{!}{\includegraphics{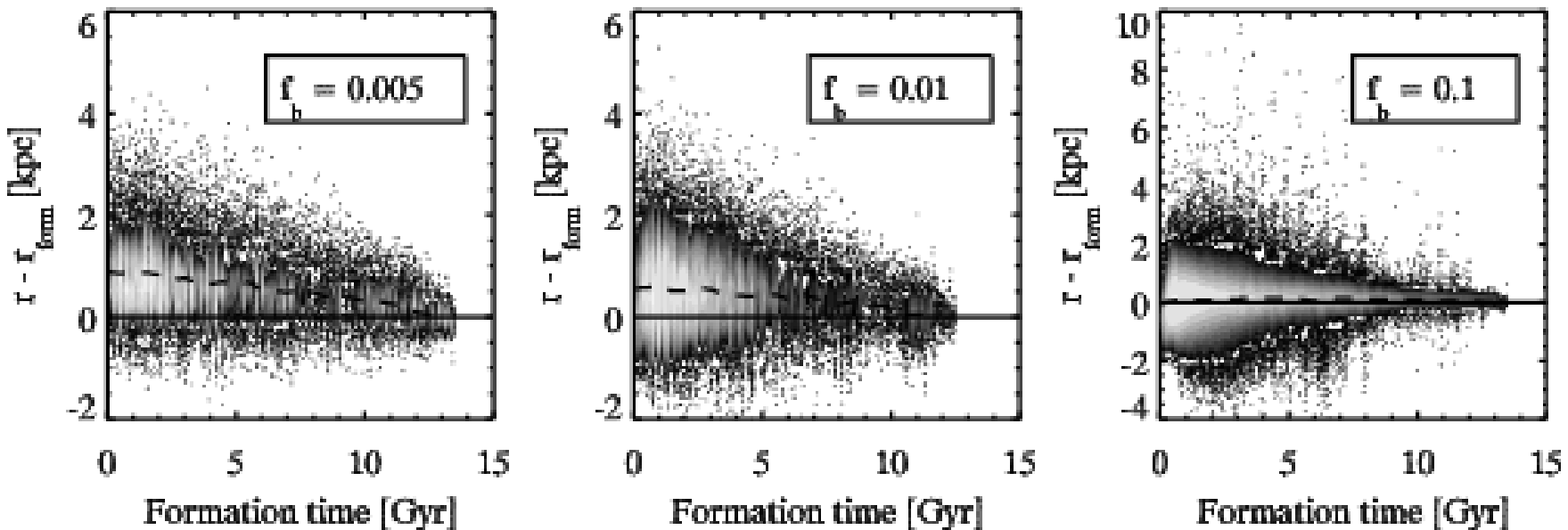}}
 \caption[Formation Time vs. Movement]{ The same as Figure \ref{fig:kmsmigration} except for the models with $v_{200}$ = 35 km s$^{-1}$ and varying $f_b$.}
\label{fig:bfmigration} 
\end{figure*}

While the location of star formation largely determines the age gradient and stellar structure of the galaxies, there are some features that are due to the motion of stars after they form.  Figures \ref{fig:kmsmigration} and \ref{fig:bfmigration} highlight these differences by showing how far the stars have migrated radially.

The mean migration (dashed line) in the constant $f_b$=0.1 models shown in Figure \ref{fig:kmsmigration} shows little deviation from zero (solid line), although many individual particles move a large distance.  For the lowest mass halos, most of this movement is in the outward direction and is concurrent with star formation episodes.  

\begin{figure*}
\resizebox{18cm}{!}{\includegraphics{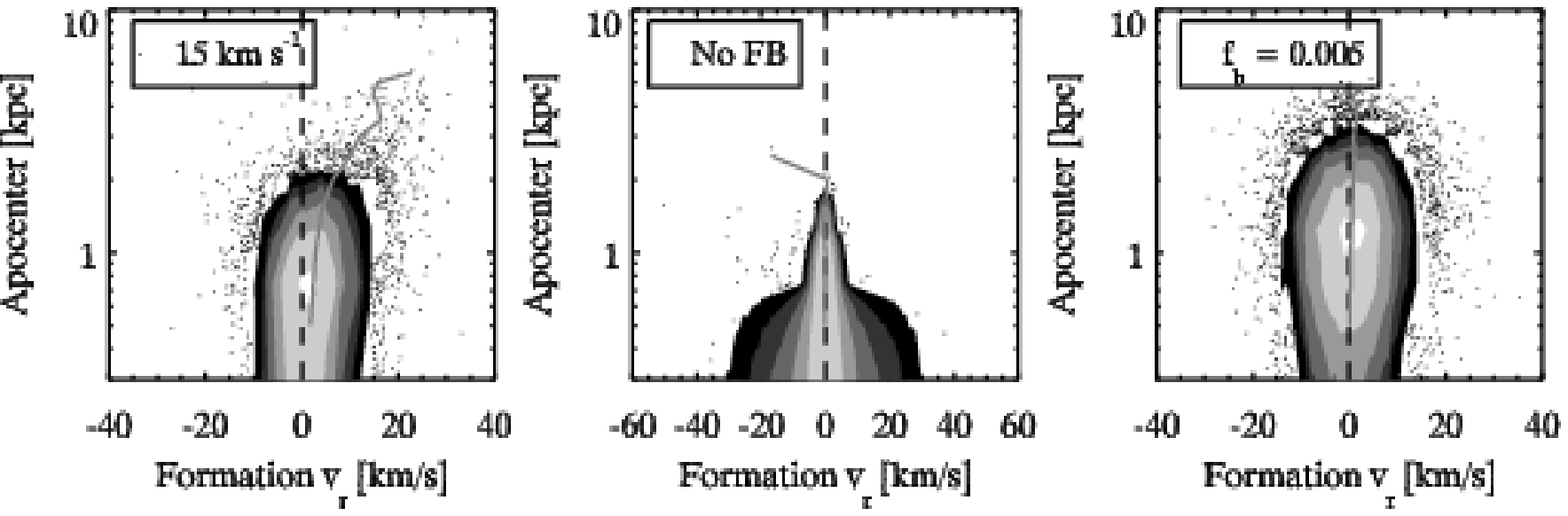}}
 \caption[Formation Radial Velocity]{ The formation radial velocity of stars compared with their apocenter (maximum radial extent) for three representative model galaxies.  Contours are linearly scaled to the density of star particles in the 1 km s\inv $\times$ 0.5 kpc size bins.  A dashed line is drawn at zero velocity to demarcate outward going forming stars from inward moving stars.  The mean formation velocity for each apocenter is plotted as the gray line.
The left panel shows the 15 km s$^{-1}$ model with only 10$^5$ gas particles.  The center panel shows what happens when the same 10$^5$ particle 35 km s$^{-1}$ model is run without supernova feedback.  The right panel shows the plot for the 35 km s$^{-1}$, $f_b$=0.5\% model run with 10$^5$ particles.  We note that the 15 km s$^{-1}$ model run with feedback shows the largest fraction of stars formed with positive outward velocities (to the right of the dotted line).  The 35 km s$^{-1}$ model shows slightly more stars forming with outward velocities than inward.  The supernova feedback is a constant energy and it is able to move stars the farthest out of the shallowest potential wells.}
\label{fig:velsf} 
\end{figure*}

To explore this behavior, Figure \ref{fig:velsf} shows the stellar apocenters as a function of their velocity at the time of formation. 
Compared to the vertical line at zero velocity, stars to the right formed from gas that was moving outwards, while stars to the left formed as the gas was collapsing.  

There is a noticeable trend for stars with larger apocenters to form with larger outward velocities, particularly for the lowest mass 15 km s$^{-1}$ halo.  
Physically, these halo stars formed from outward flowing gas that was accelerated by supernovae blastwaves and shocks against infalling gas.  The shock produces high densities and triggers star formation.  These stars are thus launched on radial orbits that create an extended
stellar halo and a positively skewed velocity distribution.  Conversely, few stars form during the infall phase during which shocks are weaker or absent.  No stars are observed to be created with sufficient energy to become unbound from the dwarf galaxy. Once the dwarf interacts with other satellites, some of these stars may become unbound.  

Figure \ref{fig:velsf} suggests that shocks and instabilities in the SN-driven wind may be necessary for the formation of the most extended halo stars seen in dwarfs.  To verify that the large radius, high initial velocity stars are indeed due to feedback, we have rerun the 15 km s$^{-1}$ simulation with feedback turned off.  In this simulation, no extended halo forms.  Instead, the distribution of apocenters is centered at zero radial velocity, consisting of disk stars formed from gas moving in stable circular orbits.  

The signature of halo formation due to supernova feedback is much reduced in higher mass galaxies, which form long-lived stable stellar disks.  In these cases, disk stars form on circular orbits and appear near the zero initial radial velocity line in Figure \ref{fig:velsf}.  These stars may migrate outwards through disk instabilities \citep{Roskar2008}, but are unlikely to swell into a 3-dimensional halo.  

Figure \ref{fig:velsf} also suggests that in the lowest mass galaxies, one expects a kinematic signature where the
outermost halo stars are all preferentially on radial orbits.
However, these stars can be strongly influenced by tidal effects, possibly
making this signature difficult to detect observationally.

\subsubsection{Stellar Disk Meandering}
\begin{figure}
\resizebox{9cm}{!}{\includegraphics{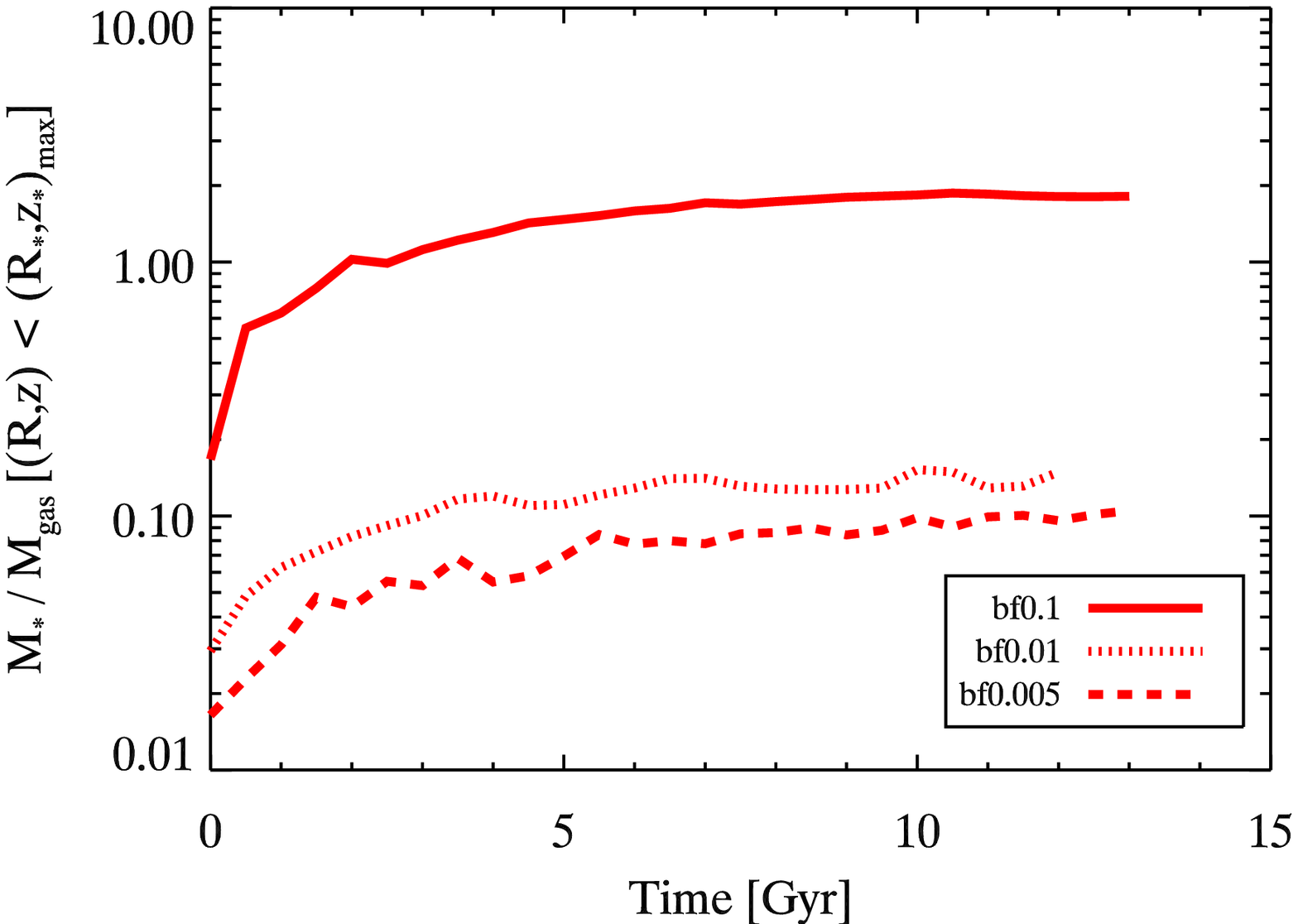}}
 \caption[Star/Gas Mass Ratio]{ The star to gas mass ratio in the stellar disk.  The ``stellar disk'' is defined using the star with the largest R and z from the center of mass of the star particles.  This definition of the disk is conservative as the gas disk stretches beyond the stellar disk in our simulations.  Even limiting the definition of the disk as we do, the larger gas mass in the $f_b$=0.5\% and $f_b$=1\% models is apparent.  This larger mass explains why the stellar disk meanders through the gas disk.}
\label{fig:gasstarmass} 
\end{figure}

Unlike in the $f_b$=10\% models, the low $f_b$ models show an offset between the mean initial and final radii as shown in Figure \ref{fig:bfmigration}.  This offset develops because while the halo potential well is deep enough that gas cools onto a disk, the gas is pressure supported and not dense enough to form stars outside the central region.  Thus, in these low $f_b$ models, the gas disk is significantly more massive than the stellar disk and dominates the disk dynamics, as shown in Figure \ref{fig:gasstarmass}.  The stellar disk then meanders in response to the dominant influence of the gas.  As the stellar disk meanders, Figure \ref{fig:bfmigration} shows that it scatters stars into the halo.  Stars that formed at the beginning of the simulation are even further from the center than those that formed at the end of the simulation.  The meandering disk therefore provides another mechanism for reinforcing the age gradient imprinted when the stars formed.  A similar mechanism is apparent in the cosmological dwarf galaxy simulation of \citet{Mashchenko2008}.
\begin{figure}
\resizebox{9cm}{!}{\includegraphics{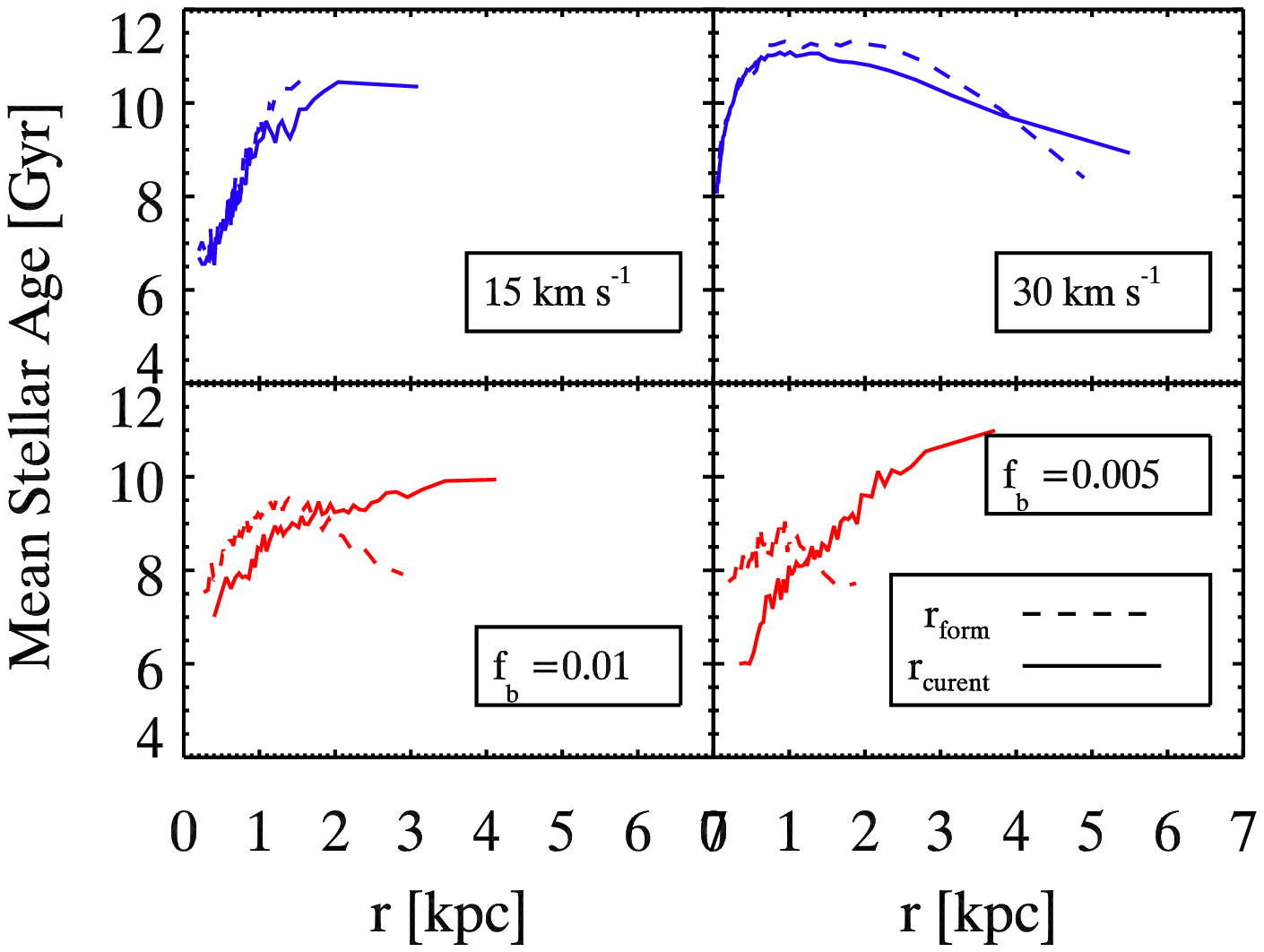}}
 \caption[Formation Age Profile]{ The mean stellar age versus formation radius.  This shows the age profile if the stars did not move at all after they formed.  The mean is taken of stars grouped together into 40 radial bins with equal numbers of particles.}
\label{fig:ageformprofile} 
\end{figure}

Figure \ref{fig:ageformprofile} shows the effect the meandering stellar disk has on the age profile.  It compares the final age profiles from Figure \ref{fig:ageprofile} with age profiles as a function of formation radius.  Figure \ref{fig:ageformprofile} thus shows the age gradient that would have been observed in the absence of stellar migration compared to the final age profile.  For the $f_b = 10\%$ simulations, the stars do not meander and their age profile is determined almost entirely by where they form.  The 35 km s$^{-1}$ models with $f_b$=0.5\% and $f_b$=1\% show declining mean stellar age at large radii at formation (dashed line), but constantly rising mean age in the present day age profile (solid line).  The sloshing of the disk is therefore capable of changing the sign of the age gradient at large radii, particularly when the baryon fraction is low.

\subsubsection{Low Temperature Cooling}
Because the disk sloshing is due primarily to high gas mass fractions, one wonders if such gas rich disks would be present if we had not neglected low temperature cooling (below $10^4$ K).  Our neglect of molecular and metal line cooling leads to an artificial temperature floor that supports a large low density gaseous disk, as explained in \citet{Kaufmann2007}.  However, low mass galaxies contain few metals, so low temperature cooling should be suppressed making our neglect of these processes less important than for massive galaxies.  Low metallicity also suppresses molecular cooling, for two reasons.  First, the formation of molecules is inefficient when dust grains are absent, greatly reducing the mass of molecular gas when the metallicity is low.  Second, molecular cooling is dominated by emission lines from metal-bearing molecules, leading to more than a factor of ten reduction in the cooling rates when metals are absent.  Moreover, there is clear evidence that even with atomic cooling alone, our simulations contain gas that is cold and dense enough to form stars.  Observationally, extended HI disks are a normal feature of dwarf galaxies \citep{Swaters2002}.  Combined, this suggests that our neglect of low temperature cooling may not be a significant limitation in these simulations.

\subsection{Broken Density Profiles}
\label{sec:bdps}
The density profile of the high resolution 15 km s$^{-1}$ model in Figure \ref{fig:hrstarprof} shows a surprising amount of structure for a galaxy that formed in isolation without any hierarchical buildup of material.  We can relate the structure to the different regions of star formation.  The innermost volume ($r < 1$ kpc) corresponds to the radius inside which stars continually form.  The next region ($1 < r < 2$ kpc) corresponds to the volume inside which heavy star formation occurred only in the past.  Beyond 2 kpc, there is a modest increase in stellar surface density slope.  This corresponds to the region in which no star formation happened \emph{in situ}, but into which stars were ejected.  Using these different regimes of star formation, it is possible to see how a dwarf galaxy that forms in isolation can create a structured stellar distribution.

\subsection{Halo Shapes}

The shapes of the stellar halos that formed in our simulations were influenced by how they were produced.  A halo dominated by stars
that formed in supernova shocks is likely to be more spherical, whereas one
that formed from disk shrinking or sloshing is likely to be more flattened like a disk.  

\begin{figure}
\resizebox{9cm}{!}{\includegraphics{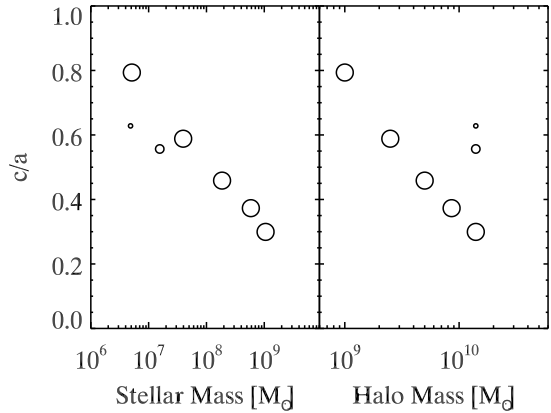}}
\caption[Halo shapes]{ Halo shape as a function of stellar mass (left panel) and halo mass (right panel) after 13.5 Gyr of evolution of the models.  The halo shape $c/a$ values come from the outermost 13\% of the stars.  The value of $c/a$ is defined in \citet{Katz91} and \citet{Debattista07} as the ratio of moment of inertia between the most minor axis and the major axis.  The size of the circles corresponds to the baryon fraction ($f_b$) of the model.  The stars on the outskirts of the smallest galaxies have the least flat shape.}
\label{fig:shapes} 
\end{figure}

Figure \ref{fig:shapes} shows the axial ratio of the outermost 13\% of stars as
a function of stellar mass (left panel) and of halo mass (right panel).
There is a general trend for systems with low stellar masses
and/or halo masses to have more spherical distributions.  The trend
with halo mass is due to the increasing importance of supernova shock-driven
halos in galaxies with shallower potential wells.  In the more massive
halos, the formation of the old stellar halo is dominated by processes that operate in disks, leading to flatter axial ratios.  

At fixed halo mass, systems
with larger baryon fractions have flatter stellar halos because their disks are more extended.  We note, however, that even the most massive high baryon fraction galaxies with the flattest halo axial ratios are still rather round in their outskirts.  At the mass scales explored in these simulations, disks exert significant pressure support, which leads to thicker star forming disks than those typical in higher mass galaxies (e.g. \citet{Kaufmann2007}).  

\section{Conclusions}
Recent, deep observations of regions surrounding Local Group Dwarfs have shown structure in their stellar populations in the form of old stellar populations at large radii and broken exponential surface density profiles.  In larger galaxies, these extended stellar halos are thought to be the result of hierarchical merging.  In our simulations, we find that even \emph{isolated} galaxies are able to form extended halos, explaining why so many dwarfs have halos, even when the expected masses of the merging sub-halos are thought to be too low to form stars.

In our simulations, three mechanisms contributed to an age gradient in the outer galaxy.  First, the envelope inside which star formation occurred contracted as the supply of gas diminished and pressure support was reduced.  Second, gas-driven disk instabilities caused bulk motion of the stellar disk over time so that stars that formed the longest ago are the farthest from the current center.  Third, some stars were ejected far out into the halo after forming in supernova driven shocks.  The ejection was somewhat more effective at early times because of the active early star formation.  

The importance of the three different mechanisms varied with mass and baryon fraction of the model.  For example, the contracting star formation envelope occurred in all of our models, but began earliest in the lowest mass models.  The higher mass models were able to initially grow disks from the ``inside-out'' using their large gas reservoir.  However, dwindling gas supplies halted the steady disk growth and star formation contracted back towards the galaxy centers, in a similar manner to the lower mass models.  

Bulk motions of the stellar disk were the most notable in the models with low baryon fraction, which replicate the effect of reionization by removing baryons from low mass halos.  Lowering the baryon fraction leads to low density gas disks that can only form stars at their center.  The low baryon fraction disks therefore remain gas rich and are more prone to instabilities that lead to sloshing of the embedded stellar disk.

The ejection of stars formed in supernova-driven shocks is most prominent in the lowest mass models.  At these mass scales, the depth of the potential well is shallow enough that supernovae can drive significant outflows, making stellar ejection more likely.

Each of these three scenarios played a role in producing a stellar distribution with a remarkable amount of structure.  Although we have not ruled out the possibility that the structure in dwarf stellar halos is the result of hierarchical merging, tidal interactions, or cold accretion along filaments, it is possible to produce extended stellar structures without merging.

\section*{Acknowledgments}

We would like to thank the anonymous referee for many helpful comments that significantly improved this paper.
We would also like to thank Rok Ro$\check{s}$kar, Peter Yoachim, Ryan Maas, Anil Seth, Chris Brook and Ben Williams, for helpful
conversations and IDL help during this project.  Adrienne Stilp kindly reran all the simulations after they were lost in a disk crash.  This work was made possible by the facilities of the Shared Hierarchical Academic Research Computing Network (SHARCNET:www.sharcnet.ca) and facilities provided by the University of Washington Student Technology Fee.  GS and TQ were supported by NSF
ITR grant PHY-0205413.  JD was partially supported by NSF CAREER AST-0238683.  TK acknowledges financial support from the Swiss National Science Foundation.

\bibliographystyle{mn2e}
\bibliography{references}

\clearpage

\end{document}